# Spectroscopic evidence for a spin and valley polarized metallic state in a non-magic-angle twisted bilayer graphene

Ya-Ning Ren[1,§], Chen Lu[2,§], Yu Zhang[1], Si-Yu Li[1], Yi-Wen Liu[1], Chao Yan[1], Zi-Han Guo[1], Cheng-Cheng Liu[2,†] , Fan Yang[2,†], and Lin He[1,3,†]

[1] Center for Advanced Quantum Studies, Department of Physics, Beijing Normal University, Beijing, 100875, People's Republic of China

[2] School of Physics, Beijing Institute of Technology, Beijing 100081, People's Republic of China

[3] State Key Laboratory of Functional Materials for Informatics, Shanghai Institute of Microsystem and Information Technology, Chinese Academy of Sciences, 865 Changning Road, Shanghai 200050, People's Republic of China

[§]These authors contributed equally to this work.

[†]Correspondence and requests for materials should be addressed to C.-C.Liu (e-mail: ccliu@bit.edu.cn), F.Y. (e-mail: yangfan_blg@bit.edu.cn) and L.H. (e-mail: helin@bnu.edu.cn).

**In the magic-angle twisted bilayer graphene (MA-TBG), strong electron-electron (*e-e*) correlations caused by the band-flattening lead to many exotic quantum phases such as superconductivity, correlated insulator, ferromagnetism, and quantum anomalous Hall effects, when its low-energy van Hove singularities (VHSs) are partially filled. Here our high-resolution scanning tunneling microscope and spectroscopy measurements demonstrate that the *e-e* correlation in a non-magic-angle TBG with a twist angle $\theta$ = 1.49 °still plays an important role in determining its electronic properties. Our most interesting observation on that sample is that when one of its VHSs is partially filled, the one associated peak in the spectrum splits into four peaks. Simultaneously, the spatial symmetry of electronic states around the split VHSs is broken by the *e-e* correlation. Our analysis based on the continuum model suggests that such a one-to-four split of the VHS originates from the formation of an interaction-driven spin-valley-polarized metallic state near the VHS, which is a new symmetry-breaking phase that has not been identified in the MA-TBG or in other systems.**

Twisted bilayer graphene (TBG) is a particularly interesting van der Waals structure with two low-energy van Hove singularities (VHSs) that can be tuned by a relative twist angle $\theta$ between the two graphene layers [1-6]. When the twist angle is tuned to near the so-called magic angle (MA) ($\theta$ ~ 1.1°), the two VHSs merge into 8 nearly flat bands around charge neutrality that are well separated from high-energy bands [7-10]. When these in-gap flat bands are partially filled, strong electron-electron (*e*-*e*) correlation drives many exotic quantum phases, including superconductivity, correlated insulator, ferromagnetism, and quantum anomalous Hall effects [11-16]. Numerous theories have been proposed to understand these interesting correlation-induced phases [17-41]. Very recently, it was also demonstrated that it is possible to realize the correlation-induced phases, which have been observed in the MA-TBG, in non-magic-angle TBG with the twist angle that is slightly larger or smaller than the magic angle [13,42]. Such a result indicates that the TBG (not limited to the MA-TBG) is a distinctively tunable platform for exploring correlated quantum states. Then, a natural question arises: can we realize new correlation-induced quantum phases in the non-magic-angle TBG, different from that observed in the MA-TBG? A common remarkable property of the TBG with small twist angles lies in that a well-defined and exactly-conserved low-energy degree of freedom, *i.e.* the valley, emerges. The electron states belonging to the valley $K$ and $K'$ would not be mutually scattered [7], leading to the valley U(1)-symmetry. Therefore, it would be very exciting if any new symmetry-breaking phase associated with the special valley degree of freedom is revealed, which is then a new state of matter undiscovered previously.

In this work, we report evidence for a correlation-induced spin and valley polarized metallic phase near the VHS in a non-magic-angle TBG with $\theta$ ~ 1.49° by using high-resolution scanning tunneling microscope and spectroscopy (STM and STS) measurements. When one of the VHSs of the 1.49° TBG is partially filled, we observe substantial broadening of both the two VHSs. Such a behavior is also observed in the MA-TBG very recently [43], indicating that the *e*-*e* interactions in the 1.49° TBG are still strong. Moreover, the *e*-*e* interactions split the partially filled VHS of the 1.49° TBG into four peaks and, simultaneously, break the spatial symmetry of electronic states around the split VHSs. Our analysis based on the continuum model for the 1.49 ° TBG indicates that the electronic correlation near the VHS will drive a spin and valley polarized metallic phase in which both the valley and spin degeneracies are lifted.

To obtain TBG with a target angle, large-area aligned graphene monolayer was grown on copper foils and the aligned graphene monolayer was cut into two pieces to fabricate TBG with a uniform twist angle [44-49] (see Methods and Supplementary Fig. S1 and Fig. S2 for the growth of aligned graphene monolayer). Then, the obtained TBG with controlled $\theta$ was transferred onto a single-crystal S-rich Cu substrate covered by an aligned graphene monolayer, as schematically shown in Fig. 1(a) (see Supplementary Fig. S3 for STM images of TBG with controlled twist angles on different Cu substrates). The twist angle between the TBG and the supporting graphene monolayer on the single-crystal Cu substrate is larger than 10 °to ensure that the TBG is electronically decoupling from the substrate [2-5,50-54]. As demonstrated previously, the topmost TBG will behave as freestanding TBG in such a case [2-5]. In the studied structure, there are S atoms, segregating from the Cu substrate, that intercalated between the graphene and the Cu substrate [55,56] (see Supplementary Fig. S4). The intercalated S atoms should also play a role in electronically isolating the TBG. Our high-magnetic-field STS measurements observe Landau quantization of the studied TBG, a characteristic feature of the 2D electron system, demonstrating explicitly that the TBG is electronically isolated from the substrate. Figure 1(b) shows a representative STM image of a 1.49 °TBG obtained in our experiment, which exhibits a moiré superlattice with the bright (dark) regions corresponding to the AA (AB/BA) stacking region. The stacking orders in the AA and AB/BA regions are further confirmed in atomic-resolution STM measurements, as shown in Fig. 1(c). We obtain a hexagonal-like lattice in the AA region, whereas we obtain a triangular lattice in the AB/BA regions.

There are two main advantages of the studied structure in the STM measurements: (1) it is convenient to obtain large-area TBG with quite uniform rotation angle; (2) the supporting substrate of the TBG is metallic. In addition, in our experimental setup, the doping of the TBG on the Cu substrate can vary slightly in different Cu terraces due to the variation of thickness of the intercalated S atoms [55,56]. This provides a new knob to engineer the novel states compared with the structure of the TBG on the hexagonal boron nitride (hBN), as studied very recently [43,57-59]. Figures 1(d) and 1(e) show representative STS spectra recorded in the 1.49 ° TBG on two different Cu terraces (labelled as Region I and Region II) and the doping in the two regions differs about 30 meV (see Supplementary Fig. S5 and Fig. S6 for more spectra). The doping of the TBG on each Cu terrace is almost the same and the slight difference of the doping of the TBG on the two Cu terraces may arise from variations of the distance between Cu substrate and graphene owning to the variation of thickness of the intercalated S atoms, as demonstrated very recently [55,56]. The spectra shown in Figs. 1(d) and 1(e) feature two low-energy sharp density-of-state (DOS) peaks, mainly localized in the AA regions of the moiré pattern, which are the two VHSs of the TBG [2-5]. Theoretically, the overlapping of energy bands of the two adjacent layers leads to other DOS peaks besides the two VHSs (see Supplementary Fig. S7), as observed in our experiment.

In order to further explore the electronic properties of the 1.49 °TBG, we carried out STS measurements in the presence of magnetic fields, as shown in Fig. 2. Figure 2(a) shows evolution of the STS spectra of the 1.49 °TBG as a function of magnetic fields. We can make two obvious observations from these spectra. The first is the appearance of high-energy Landau levels (LLs), as marked by pink arrows, with increasing the magnetic fields. Theoretically, the high-energy bands of the TBG are parabolic (Fig. 5(a)), which are quite similar as that of Bernal bilayer graphene. Therefore, these high-energy LLs arise from Landau quantization of the parabolic bands in the TBG. For simplicity, we can fit these LLs according to the Landau quantization of massive Dirac fermions in Bernal graphene bilayer [53]. The effective masses for electron and hole are roughly estimated as $(0.0168 \pm 0.0002)m_e$ and $(0.0158 \pm 0.0004)m_e$, respectively ($m_e$ is free electron mass, see Supplementary Fig. S8 for details of analysis), which agree well with that obtained in theory ~ 0.02 $m_e$. The second observation is the emergence of LLs of massless Dirac fermions between the two VHSs, as shown in Fig. 2(a). To clearly show this, we measured high-resolution STS spectra at 0.4 K as a function of magnetic fields (Fig. 2(b)) and three LLs of the massless Dirac fermions, 0 and ±1, are observed [the energy resolution is about 5 meV (1 meV) in Fig. 2(a) (Fig. 2(b)), see Supplemental materials for details]. By fitting the LLs to Landau quantization of the massless Dirac fermions, we obtain the Fermi velocity as $0.94 \times 10^5$ m/s for electrons and $0.63 \times 10^5$ m/s for holes (see Supplementary Fig. S9 for details). Obviously, the kinetic energy of the quasi-particles in the 1.49 º TBG is strongly suppressed and is more than two orders of magnitude smaller than that in graphene monolayer. As demonstrated very recently [11-16], the strongly suppression of the kinetic energy of the quasi-particles is of central role in the realization of exotic correlated states in the TBG.

In our experiment, clear evidences for *e-e* correlations are observed in the 1.49 º TBG. As shown in Figs. 1(d) and 1(e), there is an obvious interaction-enhanced energy separation between the two VHSs, $\Delta E_{VHS}$, when the chemical potential is around the charge neutrality of the 1.49 ºTBG. In region II, the right VHS is nearly half filled and the energy separation between the two VHSs is only about 55 meV. In region I, the chemical potential is in between the two VHSs and the $\Delta E_{VHS}$ increases to about 80 meV. Such a result, which has been observed very recently in the MA-TBG [43,57-59], indicates that *e-e* interactions should play an important role in determining the electronic properties of the 1.49 ºTBG.

To further explore possible many-body correlations in the 1.49 ºTBG, we carried out high-resolution STS measurements at 0.4 K (the energy resolution is 0.1 meV at 0.4 K, see Method for details). Figure 3(a) summarizes representative high-resolution STS spectra recorded in the 1.49 º TBG on different Cu terraces, showing clearly the correlation-induced splitting of the VHS when it is partially filled. Three typical high-resolution STS spectra recorded in three different regions of the 1.49 ºTBG are shown in Fig. 3(b)-3(d). The chemical potential is in between the two VHSs in the region I (Fig. 3(b)), whereas, the chemical potential crosses the right VHS in the region II (Fig. 3(c)) and region III (Fig. 3(d)). The doping in the region I and in the region II differs about 30 meV, and the doping in the region II and III differs less than 10 meV. Three notable experimental features can be observed according to the results shown in Fig. 3. The first is the interaction-enhanced $\Delta E_{VHS}$ when the chemical potential is around the charge neutrality of the 1.49 ºTBG, as also revealed in Figs. 1(d) and 1(e). The second feature observed in our experiment is the abrupt broadening of the fully occupied VHS (the left VHS) when the right VHS is partially filled. For example, the full width at half maximum (FWHM) of the fully occupied (the left) VHS is about 25 meV in the region I, however, it increases to about 43 meV (40 meV) in the region II (III) where the right VHS is partially filled. Such a phenomenon is unexpected since that there is no reason to expect a large broadening of the fully occupied bands when the occupation of the other bands is changed. Similar feature is also observed in the MA-TBG and is beyond the description of a weak coupling mean-field picture [43]. Therefore, this phenomenon is attributed to the experimental evidence that the *e*-*e* interactions play a dominant role in the MA-TBG [43]. The above two experimental features indicate that the effects of *e*-*e* interactions are also dominant in the 1.49 ºTBG even though the twist angle is larger than the magic angle by about 35%. Our result suggests that it is possible to realize strongly correlated phenomena in the TBG with a wide range of twist angle, not limited to the magic angle.

The third and maybe the most notable feature observed in our experiment is the splitting of the positive (the right) VHS into four peaks when that VHS is partially filled, as shown in Fig. 3(c) and 3(d). This one-to-four split of the VHS seems independent of the slight variations of doping between the region II and the region III. Similar feature is also clearly observed in the STM studies of the LL splittings where the *e*-*e* interactions lift both the spin and valley degeneracies and split the partial-filled LL into four peaks in the tunneling spectra [51,52]. Therefore, the one-to-four split of the VHS naturally indicates the possibility of the formation of a spin and valley polarized state in the TBG. Different from the quantum Hall isospin ferromagnetism state of graphene [51,52,60,61], no strong magnetic field is needed here and the spin-valley polarization should be purely induced by *e*-*e* interactions. According to previous studies [52,61], the valley degeneracy of the zero LL in graphene is firstly lifted, then the spin degeneracy is further removed by electron-electron interactions. Recent experiments also demonstrate that the valley degeneracy of the flat bands in the TBG will be firstly lifted at half filling [62-64]. Therefore, it is reasonable to expect that the fourfold spin-valley degeneracies of the VHS are lifted in the same way as that of the zero LL. To clearly illustrate this, we have made a schematic diagram showing how the VHS is split by Coulomb interactions in the inset of Fig. 3(c).

Very recently, correlation-induced splitting of the VHSs is also observed in the MA-TBG when the VHS is partially filled [43,57-59]. However, the VHS of the MA-TBG only splits into two peaks in the STS measurements [43,57-59]. In recent transport measurements, insulating states are observed at all integer occupancies of the eight nearly flat bands in the MA-TBG, with some cases indicating that the four-fold spin/valley degeneracies are fully lifted [14-16]. However, such spin and valley polarized states have not been detected in previous STM studies on the MA-TBG, which might be influenced by such effects as lattice relaxation, strain, local variation of the twist angle and interlayer coupling [63-66]. Furthermore, despite the peak splitting, the system should still be in a metallic phase here, as the spectra shown in Fig. 3(b) and 3(c) suggest large remaining DOS on the Fermi level. The existence of an interaction-driven spin-valley-polarized metallic phase in our experiment is further confirmed by carrying out STS maps (as summarized in Fig. 4), which directly reflect the spatial distribution of the local DOS (LDOS). In the TBG, the moiré pattern is expected to lead to spatial variation of the LDOS. For example, when the right VHS is not partially filled, the distribution of the LDOS at the VHS (Fig. 4(b)) reveals the same period and symmetry of the moiré pattern, as shown in the STM image (Fig. 4(a)). For the case that the right VHS is partially filled, the distribution of the LDOS at energies away from the VHS still exhibits the same period and symmetry of the moiré pattern but with inverted contrast, as shown in Fig. 4(c). However, the spatial distribution of the LDOS at the split VHS reveals a completely different feature comparing with that of the moiré pattern (Fig. 4(d)), indicating that the *e-e* interactions generate a new symmetry breaking phase when the VHS is partially filled.

In our experiment, the filling of the 1.49 °TBG can also be changed by magnetic fields. As shown in Fig. 2(b), more and more states are condensed into the LLs of the massless Dirac fermions with increasing the magnetic fields, which, consequently, alters the filling of the right VHS. In our experiment, the filling of the right VHS decreases gradually with increasing the magnetic fields for $B > 6$ T. As a consequence, both the broadening of the fully filled VHS (the left VHS) and the splitting of the partially filled VHS (the right VHS) decrease with increasing the magnetic fields. In Fig. 2(c) we summarize the FWHM of the fully filled VHS as a function of the magnetic fields. Obviously, it decreases for $B > 6$ T when the filling of the right VHS begins to decrease. According to our experiment, the splitting of the partially filled VHS also depends sensitively on the filling (Fig. 2(b)). When the right VHS is slightly filled, for example, at $B = 11$ T, the right VHS only splits into two peaks rather than four peaks (see Supplementary Figs. S10 for more data obtained in another region of the 1.49 $^{o}$ TBG). These results indicate that the exotic correlated states observed in the TBG should depend sensitively on the filling of the nearly flat bands, as revealed recently in transport measurements [11-16].

The splitting of the nearly half-occupied VHS into four peaks can be understood from the spin-valley-polarized order induced by the VHS under *e-e* interactions. To study such order, we start from the continuum-theory band structure [7] with a twist angle of 1.49$^{o}$, with proper relax parameters [19] (see Supplementary Figs. S11-S13 for details). The band structure thus obtained is shown in Fig. 5(a). Clearly, four low-energy flat bands (eight flat bands with considering the spin degeneracy) are obtained, which are well separated from the high-energy bands. The nonzero Fermi velocity at the $K_M$ points is $v_F = 1.5\times10^5$ m/s and the effective mass of the high-energy parabolic bands is estimated as about $0.02m_e$, which generally agree with that obtained in our experiment. The degeneracies of the high-symmetry points $\Gamma_M$, $M_M$, $K_M$ in the low-energy flat bands of the 1.49 $^{o}$TBG are the same as those of the MA-TBG [7,19]. The two peaks in the DOS (Fig. 5(b)), which reproduces the main feature observed in our experiment, are the two VHSs caused by the Lifshitz transition [23]. Figure 5(c) shows the Fermi surface (FS) at the VHS in the positive band. From Fig. 5(c), the FS-nesting near the VHS is weak, implying less possibility of inducing a density wave (DW) order.

When the system is doped to the VHS, the divergent DOS on the FS will lead to various electron instabilities [21,40,67-74], among which the leading one is determined by the channel in which the largest eigenvalues of the renormalized spin or charge susceptibility matrices $\chi^{(s/c)n_1\tau_1,n_2\tau_2}_{n_3\tau_3,n_4\tau_4}(\vec{q}, i\omega = 0)$ are the most divergent [23,75-81] (see the Supplementary Materials for details). For this purpose, we first calculated the bare susceptibility $\chi^{(0)n_1\tau_1,n_2\tau_2}_{n_3\tau_3,n_4\tau_4}(\vec{q},i\omega=0)$ at the positive VHS doping on a 2000 × 2000 lattice. The obtained largest eigenvalue as a function of $\vec{q}$ is shown along the high-symmetry lines in the moiré Brillouin zone in Fig. 5(d). Obviously, the highest peak locates at the $\Gamma_M$-point which would actually diverge in the thermal-dynamic limit due to the divergent DOS on the FS, implying an electron instability ordered at $\vec{Q}=0$. Furthermore, the eigenvector of the $\chi^{(0)}(\vec{q}=0,iw=0)$ matrix corresponding to the largest eigenvalue suggests an intra-valley order on the partially-occupied bands. Therefore, up to the zeroth-order perturbation for $\chi^{(s/c)}$, the induced orders include the intra-valley spin or/and valley polarization. To get more information for the induced orders, we introduce an interacting Hamiltonian which includes a dominating intra-valley scattering term with positive coefficient $J_1>0$ and a minor inter-valley scattering term with coefficient $J_2$, and perform perturbative treatment for the interactions. The combined three Feynman diagrams for the first-order perturbation of $\chi^{(s/c)n_1\tau_1,n_2\tau_2}_{n_3\tau_3,n_4\tau_4}(\vec{q},iw)$ shown in Fig. 5(e) leads to the following results. For $J_2>0$, a pure spin-polarized state is obtained on the partially-occupied band with order parameter in the form of $\Delta^{(s)}\sum_{\vec{k}\sigma}\sigma\left(c^+_{K\vec{k}\sigma}c_{K\vec{k}\sigma}+c^+_{K'\vec{k}\sigma}c_{K'\vec{k}\sigma}\right)$, which will split the positive VHS peak into two peaks separated by $2\Delta^{(s)}$; for $J_2<0$ a pure valley-polarized state with order parameter $\Delta^{(v)}\sum_{\vec{k}\sigma}\left(c^+_{K\vec{k}\sigma}c_{K\vec{k}\sigma}-c^+_{K'\vec{k}\sigma}c_{K'\vec{k}\sigma}\right)$ will be mixed with a locked-spin-valley-polarized state with $\Delta^{(sv)}\sum_{\vec{k}\sigma}\sigma\left(c^+_{K\vec{k}\sigma}c_{K\vec{k}\sigma}-c^+_{K'\vec{k}\sigma}c_{K'\vec{k}\sigma}\right)$ on the partially-occupied band, which will split the positive VHS into four peaks separated relatively as $\pm\Delta^{(v)}\pm\Delta^{(sv)}$. In the latter case, the electron states represented by each split peak are characterized by a definite spin and valley polarization. This is well consistent with the experimental result. We further carry out a mean-field study for the case with $J_2<0$ (see Supplementary Materials for details), which yields that the two energetically-minimized ground states are just the above obtained two valley-polarized states. The energy difference between the two valley-polarized states is very small (~$10^{-6}$ meV per unit cell) and therefore they can easily be mixed under perturbations caused by impurities or the substrate. Such mixing will lead to the unique STS spectrum consistent with our experiments.

Although the spin-valley polarized state obtained here for the $1.49^{\circ}$ TBG is induced by the VHS under electron-electron interactions, it differs from the inter-valley spin density wave (SDW) or charge density wave (CDW) states proposed for the MA-TBG [19,23] due to their different situations of the FS-nesting. In the MA-TBG, the FSs from the valley K and K' are mutually well nested (see Supplementary Materials for details). This leads to the inter-valley SDW or CDW orders with finite wave vectors, which will generally gap or suppress the VHS peak through band folding and hybridization, leading to a correlated insulator phase. However, the FSs shown here in Fig. 5(c) are not obviously nested by any vector, which leads to an intra-valley order with $\vec{Q} = 0$. Such order will generally split instead of gap the VHS peak, leaving the system metallic. The mechanism for the spin-valley polarization here is similar with the Stoner's criterion, with the only difference lying in that the extra valley degree of freedom allows for the valley polarization under attractive inter-valley interaction. We note that the VHS here belong to the type II VHS [82], which would possibly induce triplet superconductivity upon doping in the absence of FS-nesting. We leave such topic for future study.

In summary, we report evidence for strongly correlation in the 1.49 °TBG. When one VHS of the TBG is partially filled, we observe the broadening of the fully occupied VHS and, more importantly, the partially filled VHS splits into four peaks, attributing to the realization of a spin and valley polarized state. Our result indicates that it is possible to realize strongly correlated phases in TBG with a wide range of twist angle and it is also possible to realize new correlation-induced quantum phases, differing from that observed in the magic-angle TBG, in the non-magic-angle TBG.

**Method:**

**Samples Growth:** The aligned graphene monolayer was synthesized on a 25- micron-thick Cu foil (purchased from Alfa Aesar.) via a traditional low-pressure chemical vapor deposition (LPCVD) method. As shown in Fig. S1, the Cu foil was firstly heated from room temperature to 1035 ℃ in 30 min with 100 sccm (standard cubic centimeter per minute) $H_2$ flow and 50 sccm Ar flow in a horizontal tube furnace (Xiamen G-CVD system). Secondly, the Cu foil was annealed at 1035 ℃ for 12h in the same gas environment. Through long time annealing, large Cu(111) grains can be obtained. Then 5 sccm $CH_4$ was introduced to the system as carbon source for the growth of graphene, and the growth time was 30 min. In the process, graphene can follow the underlying Cu lattice's orientation. So the large-scale and spatial uniform graphene monolayer was obtained. Finally, the sample was quickly cooled down to room temperature.

By reducing the growth time of graphene, we can obtain the samples that are not completely covered with graphene. Because the copper covered by graphene is hard to be oxidized, different contrast between oxidized and unoxidized areas can be easily observed under optical microscope. Figure S2(a) shows the optical image of the sample and we can clearly observe the aligned graphene on the whole substrate. When the aligned graphene islands continue to grow, they will seamlessly splice to eliminate the grain boundaries. Then, we can obtain large-area graphene with the uniform orientation.

**The transfer of graphene sheet:** The as-grown graphene film was transferred onto a single-crystal Cu substrate by the PMMA-assisted method, as shown in Fig. 1(a). First of all, large-scale aligned monolayer graphene was synthesized on Cu foil by a traditional low-pressure chemical vapor deposition method, as shown in Fig. S1. Then, the aligned graphene monolayer was cut into two pieces and spin-coated polymethyl methacrylate (PMMA) film was transferred onto one of the pieces at 180 ℃ for 3 min. Next, the sample was immersed into the ammonium persulfate solution to etch the copper away, so the PMMA/graphene film was detached from the Cu foil. After that, we used dilute hydrochloric to clean it for hours and washed it by ultra-high purity water, then transferred PMMA/graphene membrane onto another piece with a uniform twist angle before dried in air for hours. After obtaining two layers of graphene with a target twist angle, we repeated the previous steps and transferred the TBG onto a single-crystal Cu substrate covered by an aligned graphene monolayer. Finally, we obtained TBG with controlled $\theta$ on a single-crystal Cu substrate after removing the PMMA by annealing treatment in vacuum at 550 ℃ for 6 hours. In this way, we can obtain large-area TBG with a specific rotation angle. Fig. S2(b) shows a representative 200×200 $nm^2$ STM image of one sample. Obviously, the rotation angle is quite uniform in our sample.

**STM and STS measurement.** The STM system was an ultrahigh vacuum single-probe scanning probe microscope USM-1300 from UNISOKU with magnetic fields up to 15 T. All STM images were taken in a constant-current scanning mode. The STM tips were obtained by chemical etching from a wire of Pt(80%) Ir(20%) alloys. Lateral dimensions observed in the STM images were calibrated using a standard graphene lattice as well as a Si (111)-(7×7) lattice and the STS spectra were calibrated using a Ag (111) surface. The STS spectrum, i.e., the *dI/dV-V* curve, was carried out with a standard lock-in technique by applying alternating current modulation of the bias voltage of 5 mV (793 Hz) to the tunneling bias.

The energy resolution of our experiment is influenced by both the temperature (~$3k_BT$, where $k_B$ is the Boltzmann constant) and the oscillation of bias voltage added in quadrature. The STS spectra at 4.2Kwere carried out with lock-in oscillation of 5 mV applying to the tunneling bias, which limits the energy resolution to about 5 meV. In order to acquire the spectra with high spectroscopic resolution, we lower the oscillation of lock-in bias voltage to 0.5 mV by using a 1/10 bias divider during the STS measurement at 4.2K (0.4K), where the energy resolution of high-resolution spectra is improved to about 1 meV (0.5meV).

## Acknowledgements

This work was supported by the National Natural Science Foundation of China (Grant Nos. 11974050, 11674029). L.H. also acknowledges support from the National Program for Support of Top-notch Young Professionals, support from "the Fundamental Research Funds for the Central Universities”, and support from “Chang Jiang Scholars Program”.

# Figures

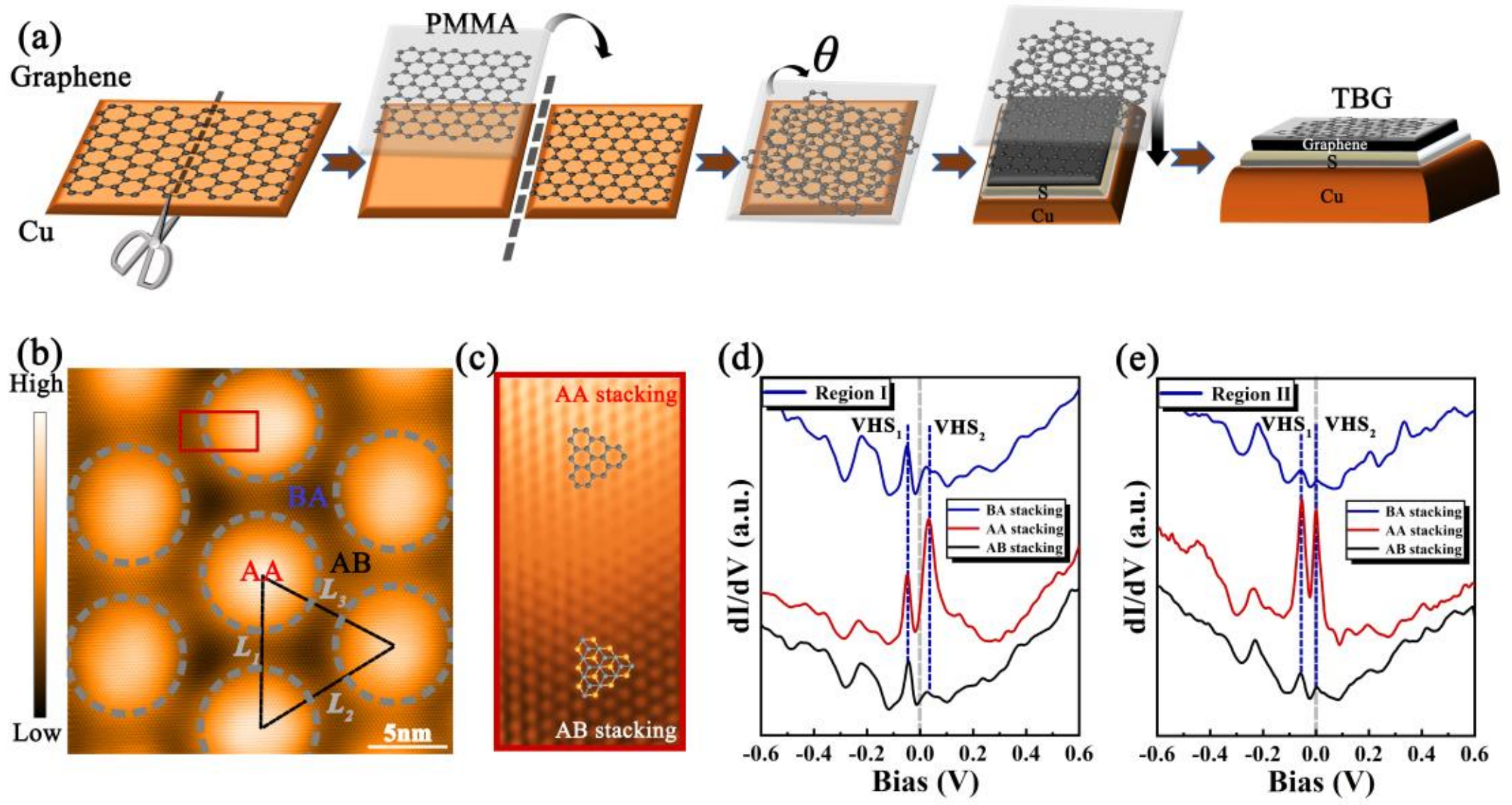

**FIG. 1.** Fabrication and STM characterization of a 1.49 °TBG. **(a)** Schematic images showing the fabrication of TBG with controlled twist angle. Aligned graphene monolayer was grown on a copper foil and the as-grown sample was cut into two flakes. Poly (methyl methacrylate) (PMMA) was spin-coated on one of the flakes and the graphene sheet was transferred onto the other one after that the copper foil was etched by ammonium persulfate. Finally, the TGB with controlled twist angle was transferred onto single-crystal S-rich Cu covered by graphene monolayer. **(b)** A 25 ×25 $nm^2$ STM topographic image ($V_{sample}$ = 60 mV, $I$ = 300 pA) showing a TBG region with twist angle θ ~ 1.49°. The periods of moiré lattice in three directions are almost the same with $L_1 \approx L_2 \approx L_3$ = 9.48 ±0.10 nm. **(c)** The atomic-resolution STM image of red rectangle area in Fig.1(b), where the AB/BA stacking regions display triangular graphene lattices, and the AA stacking regions exhibit hexagonal graphene lattices. **(d) and (e)** Typical *dI/dV* spectra recorded at region I and II with different doping of the TBG (a.u. = arbitrary units.). The two peaks around the Fermi level are the two VHSs of the TBG.

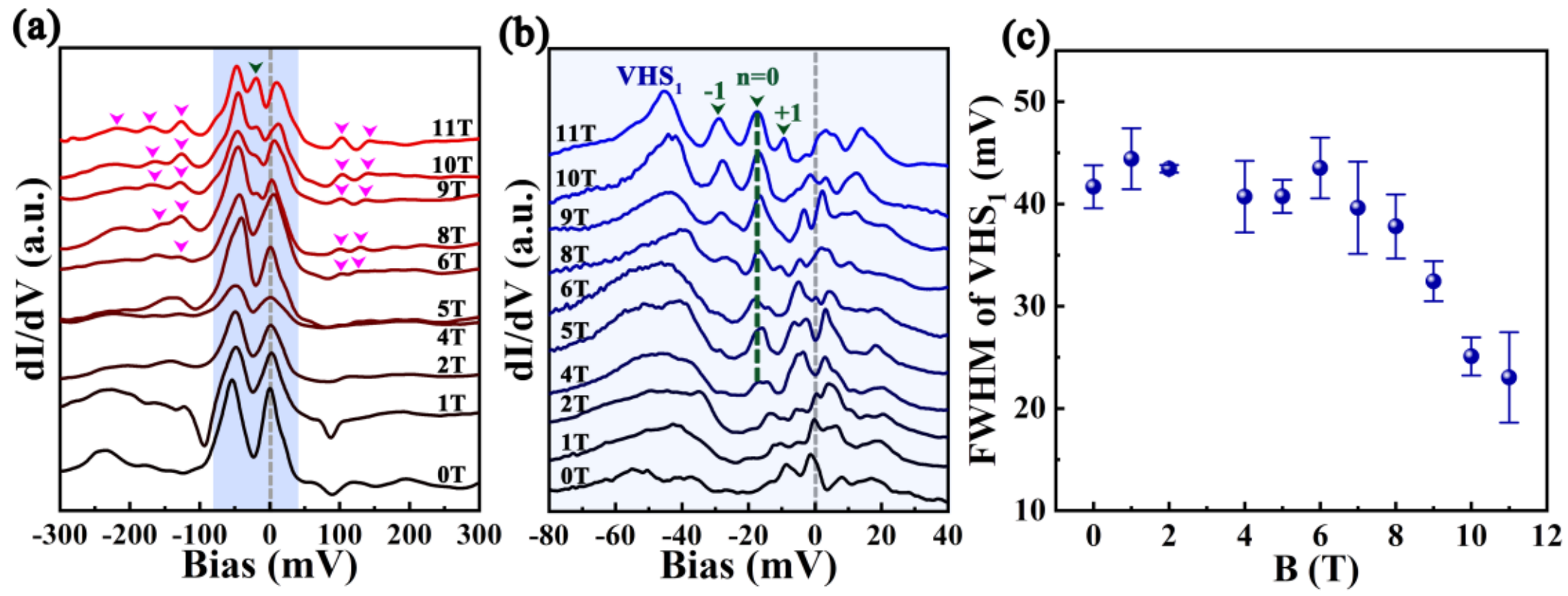

**FIG. 2.** STS spectra of the 1.49 ºTBG as a function of magnetic fields. **(a)** STS spectra (with 5 meV in energy resolution) as a function of magnetic fields *B*. The high-energy LLs, which arise from the high-energy parabolic-like bands, are marked with pink arrows. The peak between the two VHSs is the zero-LL of massless Dirac fermions. **(b)** High-resolution STS spectra (1 meV in energy resolution) measured in different magnetic fields. The measured energy region corresponds to the grey region in panel (a). The LLs of the massless Dirac fermions in the TBG are labeled. **(c)** The FWHM of the fully occupied VHS (the left VHS) as a function of magnetic fields.

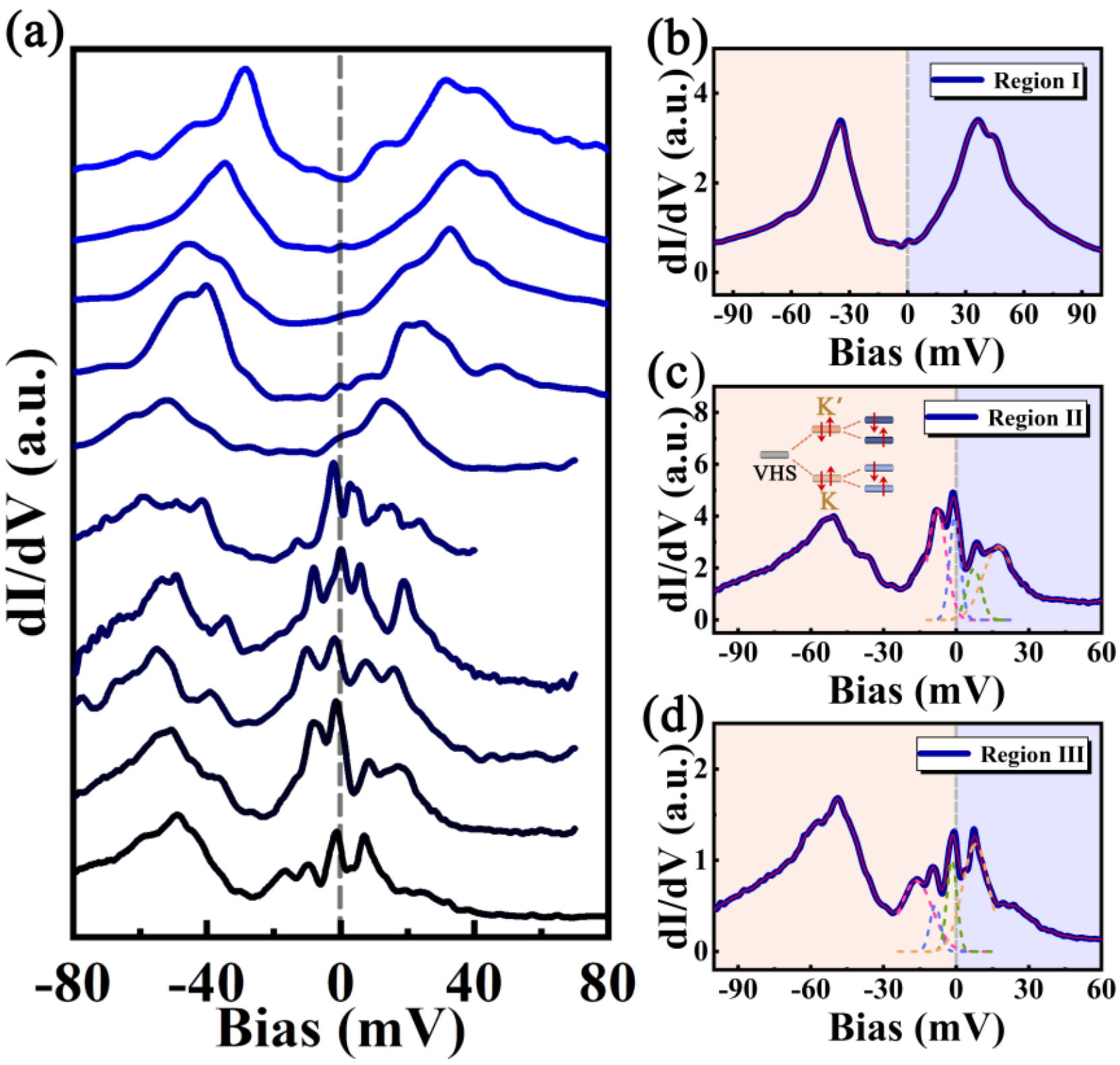

**FIG. 3.** High-resolution tunneling spectra of the 1.49 ºTBG with different fillings. **(a)** The STS spectra recorded in the AA region of the 1.49 ºTBG with different fillings at 0.4 K. Obviously, the features of the spectra depend sensitively on the filling of the right VHS. **(b)-(d)** Three representative spectra in different regions of the TBG with different fillings. In panel (b), the Fermi level is in between the two VHSs. In panels (c) and (d), the right VHS is partially filled and the occupations of the right VHS in the region II and III are slightly different. The dashed curves are the fitting of experimental data, showing that the partial filled VHS splits into four peaks. Inset in panel (c) is the schematic of the energy level structure of the VHS, showing the lifting of valley degeneracy followed by the lifting of spin degeneracy.

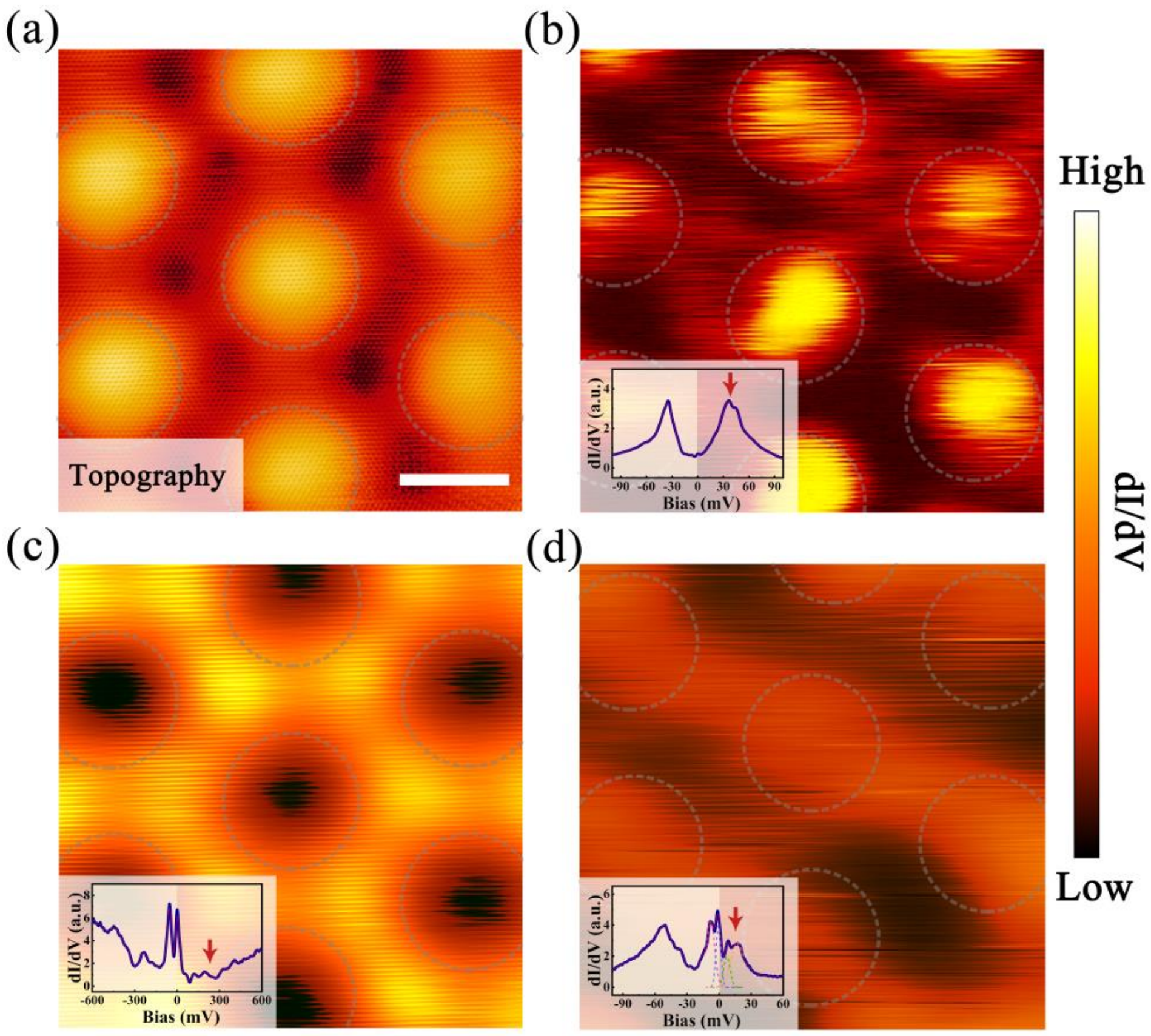

**FIG. 4.** The STS maps of the 1.49 ºTBG with different fillings. **(a)** A STM topography of the 1.49 ºTBG ($V_{sample}$ = 48 mV, $I$ = 600 pA). The dashed circles mark the period and circular symmetry of the moir épattern. **(b)** A STS map taken at 48 meV, corresponding to the energy of the unfilled VHS (as marked by the red arrow), when the Fermi level is around the charge neutrality point of the TBG. **(c)** A STS map taken at 225 meV (as marked by the red arrow), which is away from the energy of the VHSs, when the right VHS is partially filled. **(d)** A STS map taken at 20 meV (as marked by the red arrow), the energy of one of the split peaks in the VHSs, when the right VHS is partially filled. The scale bar is 5 nm for all plots.

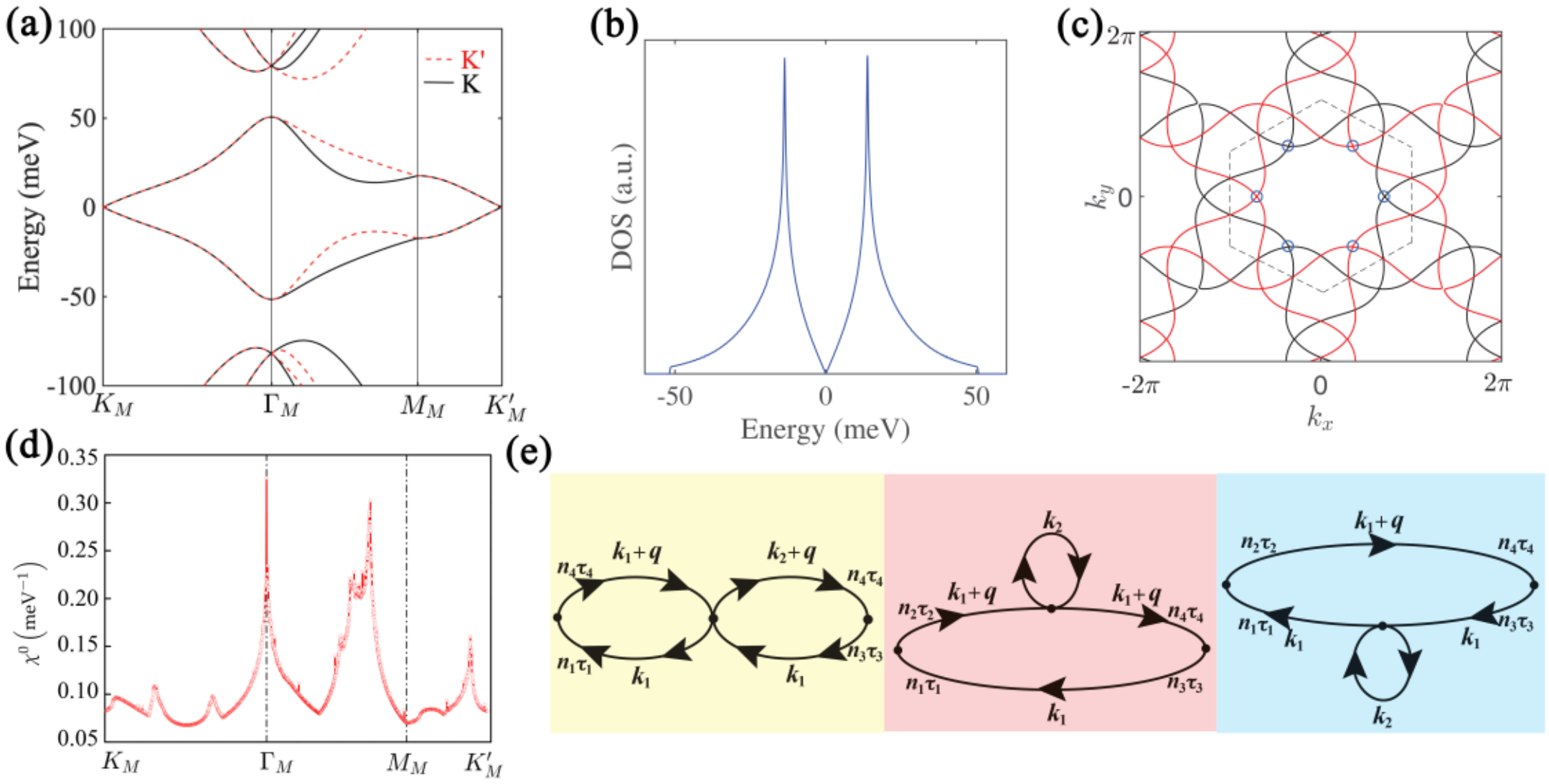

**FIG. 5.** Theoretical analysis for the VHS-induced instabilities in the 1.49° TBG. **(a)** The continuum-theory-based band structure of the 1.49 °TBG. **(b)** Density of states of the 1.49 °TBG. **(c)** FS of the 1.49 °TBG for the VHS doping. The black solid line and red dashed (solid) line in panel (a) ((c)) stand for those bands from K and K' valley, respectively. The dashed hexagon in panel (c) represents for the moiré Brillouin zone. The small blue circles on the FS denote the Lifshitz-transition points. **(d)** Distribution of the largest eigenvalue of the bare susceptibility matrix along the high-symmetry lines for the VHS doping. **(e)** The three Feyman's diagrams for the first-order perturbation of the susceptibilities.